\documentclass{emulateapj}

\usepackage{graphicx}
\usepackage{wrapfig}
\usepackage{graphics}
\usepackage{epsfig}
\usepackage{amsmath}
\usepackage{amssymb}
\usepackage{multirow}

\begin{document}

\title{Mid-Infrared High-Contrast Imaging of HD~114174~B : An Apparent Age Discrepancy in a ``Sirius-Like'' Binary System}
\author{Christopher T. Matthews\altaffilmark{1}, Justin R. Crepp\altaffilmark{1}, Andrew Skemer\altaffilmark{2}, Philip M. Hinz\altaffilmark{2}, Alexandros Gianninas\altaffilmark{3}, Mukremin Kilic\altaffilmark{3}, Michael Skrutskie\altaffilmark{4}, Vanessa P.  Bailey\altaffilmark{2}, Denis Defrere\altaffilmark{2}, Jarron Leisenring\altaffilmark{2}, Simone Esposito\altaffilmark{5}, Alfio Puglisi\altaffilmark{5} }
\altaffiltext{1}{Department of Physics, University of Notre Dame, 225 Nieuwland Science Hall, Notre Dame, IN 46556 }
\altaffiltext{2}{Department of Astronomy, University of Arizona, 993 N. Cherry Ave,  Tucson, AZ 85721}
\altaffiltext{3}{Department of Physics and Astronomy, University of Oklahoma, Norman, OK 73019}
\altaffiltext{4}{Department of Astronomy, University of Virginia, Charlottesville, VA 22904}
\altaffiltext{5}{Istituto Nazionale di Astrofisica, Osservatorio Astrofisico di Arcetri Largo E. Fermi 550125 Firenze, Italy }

\begin{abstract}

We present new observations of the faint ``Sirius-like" companion discovered to orbit HD 114174. Previous attempts to image HD~114174~B at mid-infrared wavelengths using NIRC2 at Keck have resulted in a non-detection. Our new L'-band observations taken with the Large Binocular Telescope and LMIRCam recover the companion ($\Delta L$ = 10.15 $\pm$ 0.15 mag, $\rho$ = 0.675'' $\pm$ 0.016'') with a high signal-to-noise ratio (10 $\sigma$). This measurement represents the deepest L' high-contrast imaging detection at sub-arcsecond separations to date, including extrasolar planets. We confirm that HD~114174~B has near-infrared colors consistent with the interpretation of a cool white dwarf ($J-L'$ = 0.76 $\pm$ 0.19 mag, $K-L'$ = 0.64 $\pm$ 0.20). New model fits to the object's spectral energy distribution indicate a temperature  $T_{\rm eff}$ = 4260 $\pm$ 360 K, surface gravity log g = 7.94 $\pm$ 0.03, a cooling age t$_{c} \approx$ 7.8 Gyr, and mass $M$ = 0.54 $\pm$ 0.01 $M_{\odot}$.  We find that the cooling age given by theoretical atmospheric models do not agree with the age of HD~114174~A derived from both isochronological and gyrochronological analyses.  We speculate on possible scenarios to explain the apparent age discrepancy between the primary and secondary.  HD~114174~B is a nearby benchmark white dwarf that will ultimately enable a dynamical mass estimate through continued Doppler and astrometric monitoring. Efforts to characterize its physical properties in detail will test theoretical atmospheric models and improve our understanding of white dwarf evolution, cooling, and progenitor masses.
\end{abstract}

\maketitle

\section{ Introduction}
Spatially resolved binaries provide a unique opportunity to study the composition, age, and evolution of compact objects by comparing their properties to those of the primary star. In particular, white dwarf (WD) companions orbiting nearby stars, such as Sirius and Procyon, lend themselves to detailed analysis, particularly when the metallicity and evolutionary state of the primary can be determined. Through gravitational interactions, binaries also permit calculation of a dynamical mass ({Boden}, {Torres}, \& {Latham} 2006). Model independent masses may then be used to test and calibrate theoretical WD cooling models, atmospheric models, and calculations of initial-to-final mass ratios ({Isern} {et~al.} 2007).  Directly imaged WD companions orbiting main-sequence stars (spectral type K or earlier) are, however, scarce, constituting  $\sim$8\% of the local sample of 132 WDs discovered within 20 pc of the Sun ({Holberg} 2009).  Presently only 5 ``Sirius-like systems'' (Sirius, Procyon, 40 Eridani, HR 5692, and 56 Perseus) have fully characterized orbits ({Holberg} {et~al.} 2013).

A new Sirius-like system, HD~114174, was recently discovered as part of the TRENDS high contrast imaging program ({Crepp} {et~al.} 2012b, 2013).  The primary star, HD~114174~A (Table \ref{Primary}), shows a long term radial velocity (RV) acceleration of $\sim 61\; \rm m~s^{-1}yr^{-1}$ over a 16 year time baseline. The companion's flux and neutral color, $J - K = 0.12\pm0.16$, originally led {Crepp} {et~al.} (2013) to classify the object as an $\approx$T3 brown dwarf.  However, a lower limit of 0.26$\pm$0.01 M$_{\odot}$ on the companion's mass derived from a combination of RV data and imaging led them to re-evaluate HD 114174 B as a compact object.

Initial characterization of HD 114174 B relied on Keck adaptive optics (AO) observations ({Wizinowich} {et~al.} 2000) taken in the J  and K bands using NIRC2 (instrument PI: Keith Matthews).  {Crepp} {et~al.} 2013 attempted to recover the companion in the  L'-band at Keck, but were only able to place an upper limit on its brightness based on a non-detection.  Their photometric observations were used along with white dwarf atmospheric models to characterize the newly found companion.  Since only two data points and one upper limit were available, the resulting effective temperature ($T_{\rm eff} = 8100 \pm 4000$ K) was poorly constrained, leaving the physical parameters of the WD uncertain.

\begin{figure}[h!]
\includegraphics[width=0.99\linewidth]{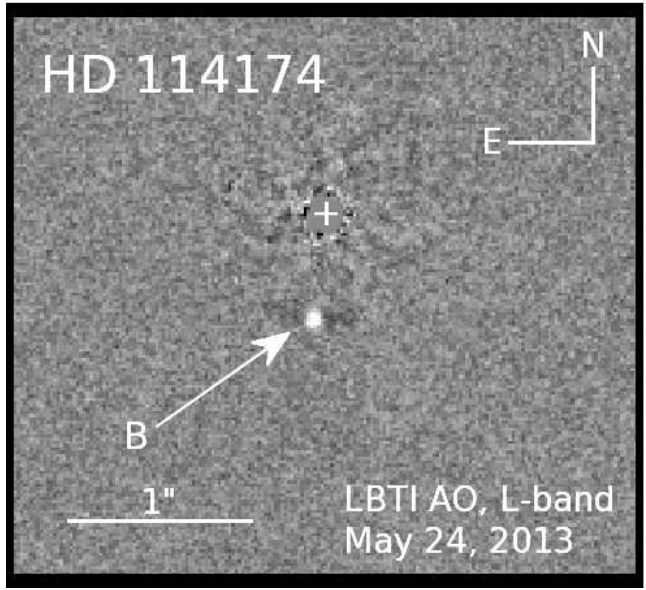}
\caption{L'-band recovery image of HD~114174~B after point-spread-function subtraction. The WD companion, HD~114174~B, is detected at an angular separation of 0.68" with a measured contrast ratio of $\Delta L' = 10.15 \pm 0.15$ at $10\sigma$. This image represents the deepest high-contrast detection within 1" in the $L'$-band to date.}
\label{L-band}
\end{figure}


%

We have acquired new photometric observations of HD~114174~B in the  L'-band ($\lambda_{c} = 3.7\;\mu$m) using the Large Binocular Telescope (LBT). The LBT AO system delivers high-fidelity wavefront correction through the use of a pyramid wavefront sensor, which offers improved noise characteristics compared to a Shack-Hartmann design, and a deformable secondary mirror (672 actuators) that provides fine spatial correction while also lowering thermal background levels at mid-infrared wavelengths ({Esposito} {et~al.} 2011; {Tozzi} {et~al.} 2008; {Close} {et~al.} 2012). These design features enable the LBT AO system to achieve unprecedented contrast levels ( {Skemer} {et~al.} 2012). 

In this Letter, we present the first imaging recovery of HD~114174~B in the L'-band (Fig. \ref{L-band}). Our new observations obtained with the LBT Interferometer (LBTI; Hinz et al. 2008) and L/M-band InfraRed Camera (LMIRCam; Skrutskie et al. 2010) rank amongst the deepest high contrast imaging in this band to date, and represents the deepest inside of 1" ($\Delta L' = 10.15 \pm 0.15$ $\rho$ = 0.675'' $\pm$ 0.016''). HR 8799~b ($\Delta L' = 10.42 \pm 0.2$) , GJ 758~B ($\Delta L' = 11.3 \pm 0.14$), and GJ 504~b ($\Delta L' = 12.9 \pm 0.19$) have been detected with greater contrast, but reside at larger separations from their primary stars: 1.72", 1.82", and 2.49" respectively ({Marois} {et~al.} 2008; {Janson} {et~al.} 2011; {Kuzuhara} {et~al.} 2013).  Additional photometric measurements\footnote{We also acquired observations in the Y-band ( $\lambda_{c} = 1.02 \mu$m) at Keck, but these resulted in a non-detection} establish a more complete spectral energy distribution (SED) thus reducing uncertainty in HD~114174~B's physical properties (Table \ref{Secondary}).  In particular, our new L'-band data lowers the effective temperature by $\sim$3800 K and reduces the relative uncertainty from 49\% to 8\%.  A cooler effective temperature corresponds to an older white dwarf.  Assuming that HD~114174~A,B have co-evolved, the new cooling age of the WD companion reveals an apparent discrepancy in their ages (2.3$\sigma$).   We also provide a new astrometric data point that will assist with on-going efforts to constrain the orbit and dynamical mass of the WD companion.

\begin{table}[h!]
\begin{minipage}{.50\textwidth}  \rule[-0pt]{9cm}{0.5pt}
\caption{\label{Primary}  Properties of HD~114174~A from {Crepp} {et~al.} 2013  }
\begin{tabular*}{\textwidth}{@{\extracolsep{\fill}}cccc}
\hline
&right ascension [J2000]& 13 08 51.02 &\\ 
&declination [J2000]& +05 12 26.06 &\\ 
&B& 7.47 $\pm$ 0.01 \\ 
&V& 6.8 $\pm$ 0.001 \\ 
&R& 6.3 $\pm$ 0.02 \\ 
&I& 6.0 $\pm$ 0.01 \\ 
&Y\footnote{We estimate the Y-band magnitude of HD~114174~A by fitting a blackbody function to BVRIJHK$_{s}$ photometric measurements (Section \ref{Y-band}).} & 5.71 $\pm$ 0.01\\ 
&J& 5.613 $\pm$ 0.026 \\ 
&H& 5.312 $\pm$ 0.027 \\ 
&K$_{s}$& 5.202 $\pm$ 0.023 \\ 
&distance [pc] & 26.14$\pm$0.37 \\ 
&Mass [M$_{\odot}$]&  1.05 $\pm$ 0.05\\ 
&Radius [R$_{\odot}$]&  1.06\\ 
&Luminosity [L$_{\odot}$]&  1.13\\ 
&Age (isochonal) [Gyr]& 4.7$^{+2.3}_{-2.6}$\\ 
&Age (gyrochronological) [Gyr]& 4.0$^{+0.96}_{-1.09}$ \\ 
&$\rm[Fe/H]$& 0.07$\pm$ 0.03 \\ 
&log(g) [cm s$^{-2}$]& 4.51 $\pm$ 0.06\\ 
&$T_{\rm eff}$ [K]& 5781 $\pm$ 44\\ 
&Spectral Type & G5 IV-V\\ 
&v sin(i) [km/s]& 1.8 $\pm$ 0.5\\ \hline
\end{tabular*}
\end{minipage}
\end{table}

\section{Observations and Data Reduction}\label{Observations}
We observed HD~114174 on UT May 24, 2013 with LBTI (Hinz et al. 2008) and LMIRCam ({Skrutskie} {et~al.} 2010; {Leisenring} {et~al.} 2012). While LBTI is capable of imaging with both telescope dishes simultaneously, we only used the left side due to a cooling incident that temporarily disabled the right-side adaptive secondary mirror. Conditions were photometric with $\sim$0.9'' seeing.  We obtained 4,500 images with an integration time of 0.291s$\times$3 coadds each at $\lambda = $3.8 $\micron$ (L').  Shortly before the observation, the LMIRCam electronics suffered from an overheating incident (now corrected) that corrupted alternating frames.  As a result, our final data set comprises $\sim$30 minutes of on-sky data taken over an 80 minute time-frame.  We did not use a coronagraph to occult the primary star.

LBTI sits at a bent Gregorian focus with no rotator, so all data is taken in ``pupil-tracking mode'' by default, where astronomical fields rotate with changing parallactic angle.  For high-contrast imaging, this allows data to be processed by Angular Differential Imaging (ADI) ({Marois} {et~al.} 2006) and derivative algorithms such as Locally Optimized Combination of Images (LOCI) ({Lafreni{\`e}re} {et~al.} 2007; {Pueyo} {et~al.} 2012) and Principal Component Analysis (PCA) ({Soummer}, {Pueyo}, \&  {Larkin} 2012; {Amara} \& {Quanz} 2012).  Over the course of our observations, the parallactic angle changed by 33 degrees.

For each frame, we interpolated over bad pixels, nod-subtracted, removed residual column bias noise, and binned the data (2$\times$2) to further suppress bad pixels.  We then registered the images and selected the best 80\% of frames for use by cross-correlation.  Finally, we coadded the images by combining 20 frames to reduce the necessary processing for our high-contrast imaging algorithm.

We used PCA to subtract an average PSF from each image, and then modeled the variability of the residuals to remove the largest variable components.  The residual images were then rotated and combined following the basic steps of ADI.  Our combined image is shown in Figure \ref{L-band}.  HD~114174~B is detected with a signal-to-noise of 10.

\section{Photometry and Astrometry}

\subsection{ L'-band Photometry}
The images of HD~114174 A were purposefully saturated in order to bring the sky background above LMIRCam's read-noise threshold.  After completing the saturated imaging sequence, we then obtained 400$\times$0.058s exposures of HD~114174~A, which served as an unsaturated PSF calibrator for measuring the relative brightness of HD~114174~B.  Like other high-contrast imaging algorithms PCA can self-subtract flux from a faint companion and bias astrometry.  Following {Skemer} {et~al.} (2012), we calibrate this effect by subtracting scaled PSFs at the position of the companion in each raw image, before running PCA.  This procedure is repeated for different positions and brightness with a grid search to determine the astrometry and photometry of the companion. We find that HD~114174~B is 10.15$\pm$0.15 magnitudes fainter than HD~114174~A, and is separated by -6.0$\pm$1.5 pixels $\times$ -31.5$\pm$1.5 pixels from the primary in detector x-y coordinates. LMIRCam's plate scale is discussed in Section 3.3. To calibrate the photometry of HD~114174~A, we observed the nearby standard LHS 2798 immediately after observing HD 114174.  We find that LHS 2798 is 0.39$\pm$0.05 mags brighter than HD~114174~A in the L'-band.  

\begin{figure*}
\begin{center}
\includegraphics[angle=0, width = \textwidth]{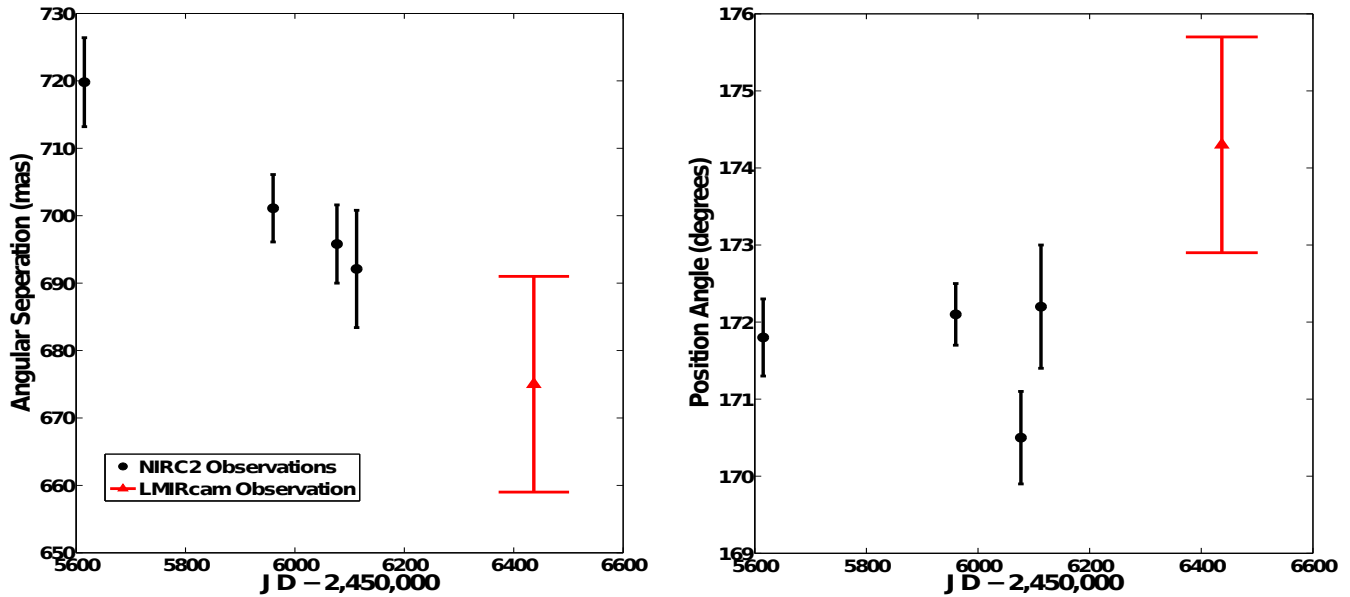}
\label{HD114174_astrometry} 
\caption{Astrometry for HD~114174~B in terms of angular separation (left) and position angle (right). Measurements from NIRC2/Keck and LMIRCam/LBT reveal systemic orbital motion.} 
\end{center}
\end{figure*}

\subsection{Y-band Upper-Limit from Non-detection}
\label{Y-band}
To further constrain the SED of HD~114174~B, we also obtained Y-band AO observations with NIRC2 at Keck on UT 2013-04-21. The star was observed at an airmass ranging between 1.1-1.2. Seeing conditions fluctuated throughout the course of the night. According to AO wavefront sensor telemetry, the seeing was $\sim$1" just prior to observations. The imaging sequence was similar to that described in Crepp et al. 2013 and also Section \ref{Observations}. We used the 600 mas diameter coronagraphic spot to occult the star.

To calibrate our Y-band photometry we also used LHS 2798 as a calibrator. Due to variability of the AO correction in the Y-band we switched between the target and calibrator star three times to account for changes in Strehl ratio. The Y-band magnitude of LHS 2798 is not recorded in the literature, so we converted BVRIJHK$_s$ apparent magnitudes from {Monet} {et~al.} 2003 and {Cutri} {et~al.} 2003 to flux measurements, and fit a blackbody spectrum to the fluxes using a grid-based search technique.  We then used our best fit to estimate the flux of LHS 2798 in the Y-band ($\lambda=1.02 \micron$).  We repeated the process for our target star HD~114174.  

We used the Aperture Photometry Tool ({Laher} {et~al.} 2012) to measure the flux ratio of HD~114174 and LHS 2798. The ratio was found to be within 8\% percent of the SED-based estimate. To determine the uncertainty in our measurement we used the $\chi^{2}$ values for our SED fits to the seven photometry points to determine the 68\% confidence interval for the effective temperature, and thus the Y-band apparent magnitude and uncertainty of HD~114174~A.  

Our Y-band observations resulted in a non-detection. We place a lower limit of $\Delta Y>8.2\pm0.7$ for the contrast ratio between parent star and companion following PSF subtraction (Table 2). Thus, HD~114174~B has an apparent magnitude $Y>14.0\pm0.7$ mag

\subsection{L'-band Astrometry}
To calibrate the $L'$ LBT astrometry measurements, we observed the Washington Double Star (WDS) Catalog\footnote{http://ad.usno.navy.mil/wds/orb6.html} standard, 70 Oph, which is classified with the WDS high grade for ephemerides accuracy ({Hartkopf}, {Mason}, \&  {Worley} 2001; {Eggenberger} {et~al.} 2008).  From three different observations dithering around the chip, we find that LMIRCam's plate-scale is $10.68\pm0.02$ mas pix$^{-1}$ and true north is counterclockwise from vertical by $0.309^{\circ}\pm0.199^{\circ}$. Applying these results to HD~114174~B, we find that the B component is separated from HD~114174~A by 0.675''$\pm$0.016'' with a position angle of $174.3^{\circ} \pm1.4^{\circ}$.

Figure 2 shows the astrometry for HD~114174~B taken over a 2.25 year baseline.  The first four points were taken with Keck/NIRC2 while the last is from LBT/LMIRCam. The combined observations self-consistently show systemic orbital motion, although the astrometry does not yet exhibit sufficient curvature to determine orbital parameters ( e.g., {Crepp} {et~al.} 2012a). Continued Doppler and astrometric observations are needed to calculate a model independent mass for this benchmark WD.

\begin{table}[h!]
\rule[-0pt]{9cm}{0.5pt}
\caption{\label{Secondary} Photometry and best fit Hydrogen model for HD~114174~B}
\begin{tabular*}{.50\textwidth}{@{\extracolsep{\fill}} cccc }
\hline 
&$\Delta Y$ & $>8.2\pm0.7$ mag  &\\
&$\Delta J$ & $10.48 \pm 0.11$ mag \\
&$\Delta K$ & $10.75 \pm 0.12$ mag \\
&$\Delta L'$ & $10.15 \pm 0.15$ mag \\
&Y& $>$14.0 $\pm$ 0.7 \\ 
&J& 16.06 $\pm$ 0.11\\ 
&K$_{s}$& 15.94 $\pm$ 0.12 \\ 
&L'&  15.30$\pm$ 0.16 \\ 
&M$_{Y}$& $>$11.85 $\pm$ 0.7 \\ 
&M$_{J}$& 13.98 $\pm$ 0.11 \\ 
&M$_{K_{s}}$& 13.85 $\pm$ 0.12 \\ 
&M$_{L}$& 13.21 $\pm$ 0.16 \\ 
\hline 
&Mass [$M_{\odot}$] & 0.54  $\pm$ 0.01 \\
&Radius [$R_{\odot}$] & 0.01301 $\pm$ 0.00019 \\
&log L/L$_{\odot}$ & -4.30 $\pm$ 0.02 \\
&Age [Gyr] & 7.77 $\pm$  0.24 \\
&log g & 7.94 $\pm$ 0.03 \\
&T$_{\rm eff}$ [K] &  4260 $\pm$ 360 \\
&M$_{v}$ &  7.90  $\pm$ 0.03 \\
&M$_{bol}$ & 15.49 $\pm$ 0.03 \\ \hline
\end{tabular*}
\end{table}

\section{Implications of YL' Photometry}

\begin{figure}
\begin{center}
\includegraphics[width=\columnwidth]{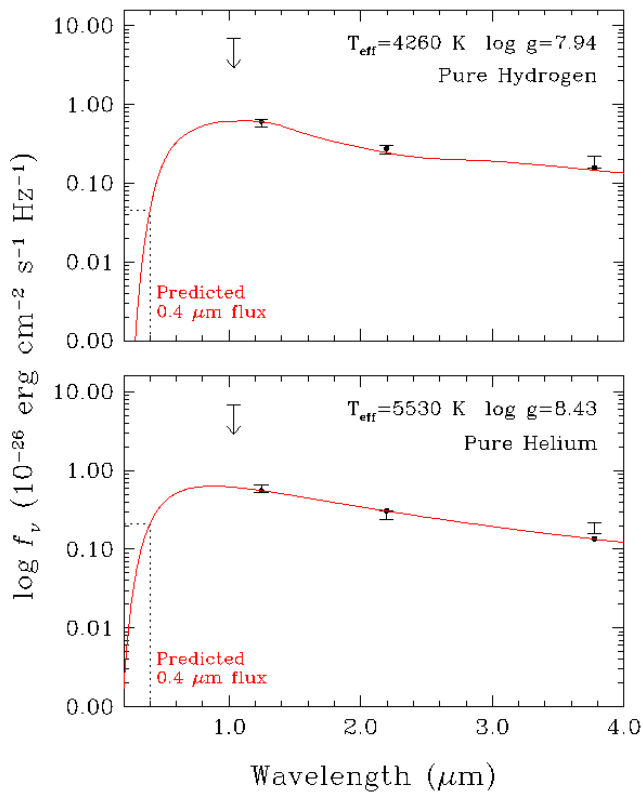}
\caption{Model fits to HD 114174 B photometry using atmospheric compositions of pure hydrogen (top) and pure helium (bottom).  Black dots correspond to synthetic flux in each filter of the best-fit model. The solid red line represents the full monochromatic SED of the best-fit model.   The dotted lines show the predicted flux at $0.4\: \micron$; note that the two models differ significantly at this short wavelength.}
\label{photo_fit}
\end{center}
\end{figure}

\subsection{WD Atmospheric Composition}
Details of our model fitting procedure are described in {Crepp} {et~al.} 2013 and {Bergeron}, {Leggett}, \&  {Ruiz} 2001. We compare models with a pure hydrogen atmosphere, a pure helium atmosphere, as well as a range of mixed atmospheres.  We used the theoretical models developed by {Tremblay} {et~al.} 2011 for our the hydrogen atmosphere, and those of {Bergeron} {et~al.} 2011 for the pure helium model atmosphere.  Mixed atmospheric models are based on the calculations of {Bergeron} \& {Leggett} 2002.

Figure \ref{photo_fit} shows the best model fits for our photometric data using pure atmospheres. While {Crepp} {et~al.} 2013 were unable to ascertain  HD~114174~B's composition, our new observations and model fits suggest that HD~114174~B has a hydrogen-rich atmosphere. Both the pure hydrogen and pure helium models produce reasonable fits to the observed photometry.  However, we obtain $\chi^{2}$=2.74 for the hydrogen solution and $\chi^{2}$=6.03 for the helium solution.  A key discriminator between the models is a precise parallax measurement from Hipparcos, $\pi$ = 38.27 $\pm$ 0.24mas, which allows for an accurate radius calculation for each effective temperature.

Accounting for the L'-band photometric observation, which greatly expands wavelength coverage, we find $T_{\rm eff}$ = 4260 $\pm$ 360 K, i.e., lower than the ($T_{\rm eff}$ = 8100 $\pm$ 4000 K) range reported in {Crepp} {et~al.} 2013. Our new Y-band upper limit does not place any significant constraint on the models but is consistent with the solution.  Table 2 displays the resulting physical properties obtained from our new hydrogen model fit for the WD companion.  Using the initial-to-final mass relations from {Williams}, {Bolte}, \&  {Koester} 2009 we find that HD~114174~B had an initial mass $M_{prog} = 1.58\pm0.17 M_{\odot}$ with the hydrogen solution, and $M_{prog} = 3.98\pm0.16 M_{\odot}$ with the helium variant.

We also consider mixed hydrogen/helium atmosphere, which have been shown to reproduce the SEDs of many cool WDs (e.g. {Bergeron}, {Leggett}, \&  {Ruiz} 2001; {Kilic} {et~al.} 2010), see however {Kowalski} \& {Saumon} 2006. 

We performed fits using a grid of mixed model atmospheres that cover a range in He abundance from log(He/H) = -2 to log(He/H) = 8. Qualitatively, the results of these fits are as good as those from pure H and He models. Examination of how $\chi^{2}$ varies as a function of log(He/H) reveals several local minima, demonstrating that there is simply too much degeneracy in the solutions from the mixed model grid to constrain the composition. Similar to the pure H and He fits, the optical/UV flux varies considerably as a function of log(He/H), reinforcing the need for optical  and/or UV data to determine HD~114174~B's composition. 

  While it is possible to acquire additional measurements in the M-band ($\lambda_{c} = 4.78\micron$) and possibly R-band ($\lambda_{c} = 0.65\micron$) using speckle-imaging ({Howell} {et~al.} 2011), our SED fits indicate that these wavelengths would provide little leverage to further understand HD~114174~B (see Figure \ref{photo_fit}).  Note that the flux at $\lambda = 0.4\: \micron$ differs significantly between the hydrogen and helium models; the hydrogen models predict a flux of 0.045 mJy whereas the helium models predict a flux of 0.209 mJy. High-resolution space-based optical/UV imaging and spectroscopy would thus permit a definitive determination of HD~114174~B's atmospheric composition. 


\subsection{An Apparent Age Discrepancy}

The cooling ages of HD~114174~B derived using our various atmospheric models do not agree with the age of the primary star.  The best fit hydrogen model yields a cooling age $t_{c} = 7.77 \pm  0.24 $ Gyr, discrepant with both the isochronological age of t = 4.7$^{+2.3}_{-2.6}$ Gyr and the more precise gyrochronological age t = 4.0$^{+0.96}_{-1.09}$ Gyr.  The helium models produce a hotter white dwarf and hence a shorter cooling age, t$_{c}$= 6.17 $\pm$ 0.03 Gyr, which differs by 2.3$\sigma$ from the gyrochronological age, and results in a $\Delta \chi ^{2} = 3.29$ worse fit to the data.

  Considering situations in which the primary and secondary are co-eval (i.e. non-capture scenarios), limited data combined with inaccuracies or systematic errors in theoretical models may produce an erroneous age for the WD. For example {Salaris} {et~al.} 2009 performed an examination of the systematic error sources that initial-to-final-mass relations for WDs. They found that the input physics for WD models was a small error term compared to the uncertainty in the stellar evolution tracks for the progenitor, but that the choice of treatment for model atmospheres, neutrino cooling, and conductive opacity can have a significant effect on the estimation of the cooling age for WDs.  

Alternatively there exist several physical explanations for the apparent age disparity.  HD~114174 may contain a faint and undetected debris disk, however given its old age and cool temperature this situation is unlikely ({Kilic} {et~al.} 2009).  HD~114174 is nearby ($\pi$ = 38.27 $\pm$ 0.24 mas)  and above the plane of the galaxy so we also do not expect significant extinction ($\Delta J < 0.02$ mags; {Schlafly} \& {Finkbeiner} (2011)).  The current separation of HD~114174~B (59.8 $\pm$ 0.4 AU) precludes any significant past mass exchange via Roche lobe overflow; however, dynamic events involving the ejection of a third body could have moved the secondary out from a much closer original orbit ({Delgado-Donate} \& {Clarke} 2008). 

  {Jeffries} \& {Stevens} 1996 have proposed that slow, massive winds from the AGB progenitor of a WD can ``spin-up'' a main sequence companion, making it appear younger.  In fact a new directly imaged wide-separation (50 AU) WD system appears to exhibit an age discrepancy similar to that of HD~114174, and provides indirect evidence for the wind accretion mechanism. {Zurlo} {et~al.} (2013) report the direct detection of a hot (18,800 K) WD around an anomalously rapidly rotating K-dwarf, HD~8049.  They argue that the unusually high rotational velocity of the primary star indicates that it must have at some point exchanged mass and angular momentum with the WD progenitor, and that this exchange is the source of the apparent age difference.  In any case further observations are clearly warranted to help refine the age estimates of HD~114174~B and place age discrepancies, if any, within the context of similar systems such as HD~8049.

\section{Summary and Discussion}
Recent advances in AO correction and wavefront sensor technology (Esposito et al. 2011) combined with infrared-optimized instrumentation available at the LBT (Hinz et al. 2008; Skrutskie et al. 2010) provide a unique platform for high-contrast imaging studies in the 3-5 $\micron$ wavelength range. We have acquired new $L' $ observations of the faint WD companion orbiting HD~114174, a purported benchmark WD discovered by the TRENDS high-contrast imaging survey ({Crepp} {et~al.} 2013). HD~114174~B was imaged previously at near-infrared wavelengths, but observations from Keck using NIRC2 resulted in a non-detection in the $L'$-band. Using LMIRCam at the LBT, we recover the companion at $10\sigma$. HD~114174~B has an apparent magnitude $L'=15.3\pm0.16$ mag and is $\Delta L'=10.15\pm0.15$ magnitudes fainter than its parent star. Our observations constitute the largest $\Delta L'$ high-contrast imaging detection at subarcsecond separations to date.

By extending the wavelength coverage for this ``Sirius-like" WD, we are able to place stronger constraints on its physical properties compared to the discovery paper. Our observations support the interpretation that HD~114174~B is a compact object with neutral infrared colors ($J - L'= 0.76 \pm 0.19$ mag, $K-L'=0.64\pm0.20$).  New model fits suggest that a pure hydrogen atmosphere is preferred over a pure helium atmosphere, although mixed atmospheres provide similar results.  Irrespective of its chemical composition, we confirm that HD~114174~B is an old and cool benchmark WD.


For plausible mixtures of hydrogen and helium, we find that the effective temperature of HD~114174~B is significantly lower than originally determined by {Crepp} {et~al.} 2013 and hence our updated cooling age is longer.  However, a longer cooling age ($t_{c} = 7.77 \pm  0.24$ Gyr) is inconsistent with the age of HD~114174~A (t=4.7$^{+2.3}_{-2.6}$ Gyr from isochrones, t = 4.0$^{+0.96}_{-1.09}$ Gyr from gyrochronology).  This disagreement is exacerbated by the lower estimated progenitor mass compared with {Crepp} {et~al.} 2013: a $M_{prog}$ = 1.58 $\pm 0.17 M_{\odot}$ star will have a substantial ($\sim$3.2 Gyr) main sequence lifetime.  This apparent discrepancy could be the result of limited photometric data combined with inadequates in WD models at near-IR wavelengths, or it may be the manifestation of a physical process involving winds from the WD progenitor that increase the spin rate of the companion star. Diffraction-limited imaging and spectroscopic measurements at optical and UV wavelengths from space using HST/STIS would provide vital leverage for characterizing HD~114174~B's atmosphere (see Fig. 3).  Such observations would eliminate the ambiguity in the WD's effective temperature, which in turn would refine HD~114174~B's cooling age, shedding light on the history and properties of this benchmark system.

\section{ACKNOWLEDGMENTS}
CTM and JRC acknowledge support from NASA origins of solar systems grant NNX13AB03G.  MK acknowledges the support of the NSF under grant AST-1312678. AS was supported by NASA Origins of Solar Systems grant NNX13AJ17G.  The LBT is an international collaboration among institutions in the United States, Italy and Germany. LBT Corporation partners are: The University of Arizona on behalf of the Arizona university system; Istituto Nazionale di Astrofisica, Italy; LBT Beteiligungsgesellschaft, Germany, representing the Max-Planck Society, the Astrophysical Institute Potsdam, and Heidelberg University; The Ohio State University, and The Research Corporation, on behalf of The University of Notre Dame, University of Minnesota and University of Virginia.The LBT Interferometer is funded by the NASA as part of its Exoplanet Exploration program.  LMIRCam is funded by the NSF through grant AST-0705296. 


\end{document}